\documentclass[preprint,pre,aps,superscriptaddress]{revtex4-1}

\usepackage{amsmath}    % need for subequations
\usepackage{graphicx}   % need for figures
\usepackage{color}      % use if color is used in text
\usepackage{subfigure}  % use for side-by-side figures
\usepackage{hyperref}   % use for hypertext links, including those to external documents and URLs
\usepackage[]{natbib}%numbers

\newcommand{\average}[1]{
  \left< #1 \right>
}
\newcommand{\MDlambda}{0.50}
\newcommand{\MDec}{-11.9}
\newcommand{\MDsigma}{2.9}
\newcommand{\MDkappaSmaller}{0.56}
\newcommand{\MDgammaZero}{4.82}
\newcommand{\MDEcut}{-25}
\newcommand{\kc}{15}

\newcommand{\picturewidth}{1\columnwidth}

\begin{document}
\title {How coupled elementary units determine the dynamics of macroscopic glass-forming systems}
\author{Christian Rehwald}
\author{Andreas Heuer}
\affiliation{Institut f\"ur Physikalische Chemie, Westf\"alische Wilhelms-Universit\"at M\"unster,
Corrensstrasse 28/30, 48149 M\"unster, Germany}
\date{\today}

\begin{abstract}
We investigate the dynamics of a binary mixture Lennard-Jones system of different system
sizes with respect to the importance of the properties of the underlying potential energy landscape
(PEL). We show that the dynamics of small systems can be very well described within the continuous
time random walk formalism, which is determined solely by PEL parameters.
Finite size
analysis shows that the diffusivity of large and small systems are very similar. This suggests that
the PEL parameters of the small system also determine the local dynamics in
large systems. The structural relaxation time, however, displays significant finite size effects.
Furthermore, using a non-equilibrium configuration of a large system, we find that causal
connections exist between close-by regions of the system.
These findings can be described by the coupled landscape model for which a macroscopic system is
described by a superposition of elementary systems each described by its PEL.
A minimum coupling is introduced which accounts for the finite size behavior.
The coupling strength, as the single adjustable parameter, becomes smaller closer to the glass
transition.
\end{abstract}

\maketitle

\section{Introduction}
Understanding the complex dynamics of glassy materials such as supercooled liquids is still a
highly controversial problem. In contrast to regular liquids the single particle dynamics in the
supercooled state become spatially and temporally correlated, which leads to pronounced
dynamical heterogeneities. Dynamical heterogeneities are a universal property of glass-forming
systems and are responsible for several phenomena like non-exponential decay of response
functions and the violation of the Stokes-Einstein relation.

In small model systems Vogel and coworkers \cite{Vogel2004} could show that particle rearrangements
are typically
localized and that their number does not depend on temperature. These
findings could also be validated for large systems by Keys \emph{et al.} \cite{Keys2011}. The
authors additionally related
the propagation of
motion to dynamical subunits which occur in a facilitation-like manner, meaning that excitations on
one scale facilitate dynamics of neighboring excitations thereby creating excitations on larger
scales. The characterization of facilitated dynamics in molecular dynamics (MD) simulations was
attempted by different authors. For example Candelier and coworkers
\cite{Candelier2010,Candelier2010a} proved the
aggregation of
cage-breaking processes into avalanches in a granular system.
Presently discussed models range between the old idea of Adam
and Gibbs \cite{Adam1965} of
a qualitative decomposition of a sample into elementary subsystems and the 
facilitation approach, where all complexity emerges by more or less complex coupling rules.
For
example the kinetically constrained spin models
were studied extensively by different authors \cite{Garrahan2002,Hedges2009}. Dynamical facilitation
can
alternatively be formulated as a coupling process. Rehwald \emph{et al.} \cite{Rehwald2010}
discovered from studying
finite
size effects, that these coupling processes determine structural relaxation properties. In shear
experiments it was demonstrated that the range of coupling 
interactions, manifested by correlated particle displacements due to stress, covers just some
particle diameters \cite{Bedorf2010,Zink2006}. Other groups report of long ranged elastic
interactions in glycerol measured by dielectric relaxation \cite{Pronin2011}.

A different approach, relating the dramatic slowing down of the dynamics at the glass transition to
the potential
energy landscape (PEL), was introduced by Goldstein \cite{Goldstein1969}. He stated that at
sufficiently
low temperatures the system resides near local minima, so called inherent structures (IS), of the
potential. While many computer simulations confirmed that the thermodynamics are indeed determined
by the distribution of ISs, the connection with the dynamics could be revealed much later. For
small systems it was possible to predict the dynamics from parameters of the underlying potential
energy landscape (PEL) \cite{Denny2003,Doliwa2003a,Souza2008,Mauro2007} with the concept of metabasins. The
dynamics
between
metabasins are closely related to the trap model \cite{Monthus1996,Dyre1995}. Many features of the
complex glassy dynamics could be derived from PEL properties. 

The situation changes when studying large systems: the PEL is no longer suitable for the prediction
of the dynamics. Here we show that the PEL formalism can be extended to large systems by
decomposing the system into elementary subsystems each of which is described by its own PEL. In
contrast to kinetically constrained spin models the elementary subunits in this approach are complex
systems, already reflecting macroscopic properties for thermodynamic observables \cite{Doliwa2003}
or self diffusion properties. Since thermodynamic properties do not change significantly when
enlarging the system above a certain minimum size \cite{Doliwa2003}, and rearrangements are
localized within a
temperature independent size, we can be sure to capture the important relaxation properties in a
single elementary subsystem. Additional fluctuations and interactions can then be included into the
coupling rules between the subsystems. In terms of the complex nature of the elementary subsystem
our approach resembles the mosaic approach \cite{Lubchenko2007} where the sample is decomposed
into a mosaic of aperiodic crystals, so called entropic droplets. Nevertheless, in the mosaic
approach all coupling processes are captured by the distribution of free energy barrier heights.

This manuscript is organized as follows: First we show how the minimum system can be described by
the PEL and the continuous time random walk approach (CTRW), both analytically and numerically.
After
discussing changes of the CTRW when approaching larger system sizes we
give a short summary of the physical scenario and motivate the presence of  coupling effects. After
discussing some technical questions concerning the used variables for the comparison between model
and MD, we explicitly identify the presence of coupling processes via appropriate simulations.
These findings motivate the coupled landscape model (CLM) which will be presented in the last
section. It is shown that to a very good approximation the finite size effects of dynamic
observables are fully captured by the CLM. Finally we figure out in how far the CLM is connected to
recently discussed models.

% \begin{figure}[bt]
% \includegraphics[width=\picturewidth]
% {Figure1}
% \caption{Estimation of $\sigma_\gamma$ of the log-normal distribution for the jump rate using MB
% waiting times for different temperatures $T$. The gray region corresponds to the error bars, the
% black vertical lines mark the $\pm2\sigma$ interval of $p_{eq}(e)$.}
% \label{fig: FluctuatingRateSigma}
% \end{figure}

\section{The minimum system}

\begin{table}
\setlength{\tabcolsep}{10pt}
\begin{tabular}{c|c|c|c|c|c|c}
\toprule
$\sigma$ & $\lambda$ & $e_c$ & $\kappa$  & $\Gamma_0$ & $e_{cut}$ & $\sigma_\gamma$\\
\colrule
\MDsigma	& \MDlambda & \MDec & \MDkappaSmaller & \MDgammaZero & \MDEcut & 1.0\\
\botrule
\end{tabular} 
\caption{Thermodynamic and dynamic PEL parameters for the binary mixture Lennard-Jones system with
$N=65$ and $e_0=0$.}
\label{tab: parameter}
\end{table}

Throughout this paper we compare the model results with the binary mixture Kob-Andersen
Lennard-Jones (BMLJ) system \cite{Kob1995} applying periodic boundary conditions. Due to the small
system size we have used a slightly shorter cutoff of $r_c=1.8$ \cite{Doliwa2003a}. All MD
simulations have been performed in the NVT ensemble using a Nos\'e-Hoover thermostat. 
  
The PEL of a binary mixture of Lennard-Jones particles applying different system sizes was
studied
extensively \cite{Heuer2008}. One key result of Doliwa and Heuer \cite{Doliwa2003} is the
connection between states of the PEL, so called metabasins (MB), and the dynamics of the system. An
MB can be constructed from a given trajectory of inherent structures by removing the complete
forward-backward correlations between them. If the system is small enough, the mean waiting time
$\average{\tau(e)}$, during which the system resides in a MB, strongly depends on the energy $e$ of
the specific MB.
Of course, the system cannot be too small because otherwise massive finite size effects set in
regarding the
thermodynamics \cite{Doliwa2003a}. In the same work it has been shown, that a system size of
$N_\text{min}=65$ (BMLJ65) particles is a good choice for a minimum system. This minimum length
scale of approx. 65 particles does not display any significant temperature dependence.

In the next step one has to determine how the specific properties of the potential
energy landscape determine the dynamic properties of the liquid. Many properties will resemble the
simple trapmodel \cite{Monthus1996,Diezemann2011}. Here we summarize the most
important results. The corresponding parameters, determined from appropriate simulations
\cite{Heuer2008} are listed in Tab. \ref{tab: parameter}. Due to  improved simulation data they
slightly differ from those given in \cite{Heuer2008}.

The shape of the density of states (DOS) $G(e)$ turned out to be close to a
Gaussian $G(e)\sim\exp[-(e-e_0)/2\sigma^2]$ with an additional cutoff below $e_{cut}$.
For low $e$ the $G(e)$ decays slightly faster than expected from a Gaussian. For reasons of
simplicity this is modeled by a cutoff energy $e_{cut}$ and an additional factor
$\exp[-(e-e_{cut})^\mu]$. For analytical results this factor will be neglected. In the trap
model the escape out of a MB of energy $e$ can be described by a simple rate process with rate
$\Gamma(e)$, given by $\Gamma(e)=\Gamma_0\exp[\beta e]$. For the BMLJ65 system a slightly more
complex energy dependence is observed \cite{Heuer2008}:

\begin{align}
\frac{\Gamma(e)}{\Gamma_0}=&
  e^{-\beta V_0}
  \begin{cases}
    \exp\left[\lambda(\beta+\kappa k_\text{entro})(e-e_{c})\right]& , e<e_c\\
    1 & ,e>e_c
  \end{cases},
\end{align}
where $\Gamma_0$ defines the the overall time scale. The escape from a MB of energy $e$ is
a multi-step process. The energy of the first IS after having completed the escape process is
denoted by $e_c$. Due to percolation arguments $e_{c}$ is independent of $e$ \cite{Heuer2008}. The
energy at the final barrier is given by $V_0+e_c$, i.e. $V_0$ denotes the height of the last barrier
crossed. The relaxation is
solid-like (activated) for $e<e_{c}$, whereas it is liquid-like otherwise \cite{Doliwa2003c}. For
$e<e_c$ an additional
entropic term, involving the factor $\kappa k_\text{entro}$, has to be taken into account, where
$k_\text{entro}=(e_0-e_c)/\sigma^2$. It reflects that the number of escape paths from a MB
increases exponentially with decreasing MB energy. The limit $\kappa=1$ can be rationalized  in a
simple PEL model \cite{Brawer1984}. For reasons of simplicity we chose the energy scale such
that $e_0=0$ from now on. In the general case $\kappa<1$ plays the role of an empirical factor of
the order of one. 
If the investigated system
contains $M$ independent subsystems one gets $\lambda=1/M$. Some additional implications of the
choice
$\lambda<1$ will be discussed in the appendix.

% The entropic prefactor could in principle be temperature dependent, but since it is very
% challenging
% to determine $\kappa$ with sufficient accuracy, this remaining dependence is smaller than our
% numerical error. Therefore $\kappa$ can be assumed to be constant.

As reported in \cite{Heuer2005}, the width of the waiting time distribution at fixed energy $e$,
expressed via $S(e)=\average{\tau^2(e)}/\average{\tau(e)}^2 -1$, show deviations from an exponential
distribution which disagrees with a pure rate process. This is, however, expected if the total
system
is a superposition of subsystems \cite{Heuer2008}. This broadening results from the fact that the
total energy $e$
can be decomposed in different ways. Formally, this broadening can be expressed by a distribution
$\varpi(\gamma,e)$ of jump rates $\gamma$ at fixed energy $e$. Here we approximate the
$\gamma$-dependence of $\varpi(\gamma,e)$ by a log-normal
distribution with
variance $\sigma_\gamma$ and mean value $\mu(e)$.
Instead of an exponential distribution one gets $\omega(\tau,e)=\int\text{d}\gamma
\varpi(\gamma,e)\gamma\exp[-\gamma\tau]$ as waiting time distribution for states of energy $e$. The
moments of this distribution are related to the moments of $\gamma$ via
\begin{equation}
 \average{\tau^n(e)}_\omega = n!\frac{\exp[n^2\sigma_\gamma^2]}{\average{\gamma^n(e)}_\varpi}\quad.
\end{equation}
In the MD simulation only $\average{\tau^n(e)}$ can be calculated. Therefore one can estimate
$\sigma_\gamma$ by the relation 
\begin{equation}
\label{eq: calc sigma gamma}
 \sigma_\gamma^2 = \ln\left[\frac{\average{\tau^2(e)}_\omega}{2\average{\tau(e)}_\omega^2}\right]
\end{equation}
We calculate the right hand side of equation \eqref{eq: calc sigma gamma} from MB
trajectories and find that $\sigma_\gamma$ is approx. 1.0, corresponding to a broadening of one
order of
magnitude. It is slightly temperature
dependent and can to a good approximation be assumed to be constant in the relevant energy range.
The
distribution $\varpi(\gamma,e)$ is very narrow compared to the equilibrium
distribution of rates $p(\log\Gamma)$ (standard deviation at $T=0.5$ is approx. 4.4).
We note in passing, that for $\sigma_\Gamma>0$ the energy of a state is
no longer solely sufficient to describe the specific rate $\gamma$ of a MB.

\begin{figure}[tb]
\includegraphics[width=\picturewidth]
{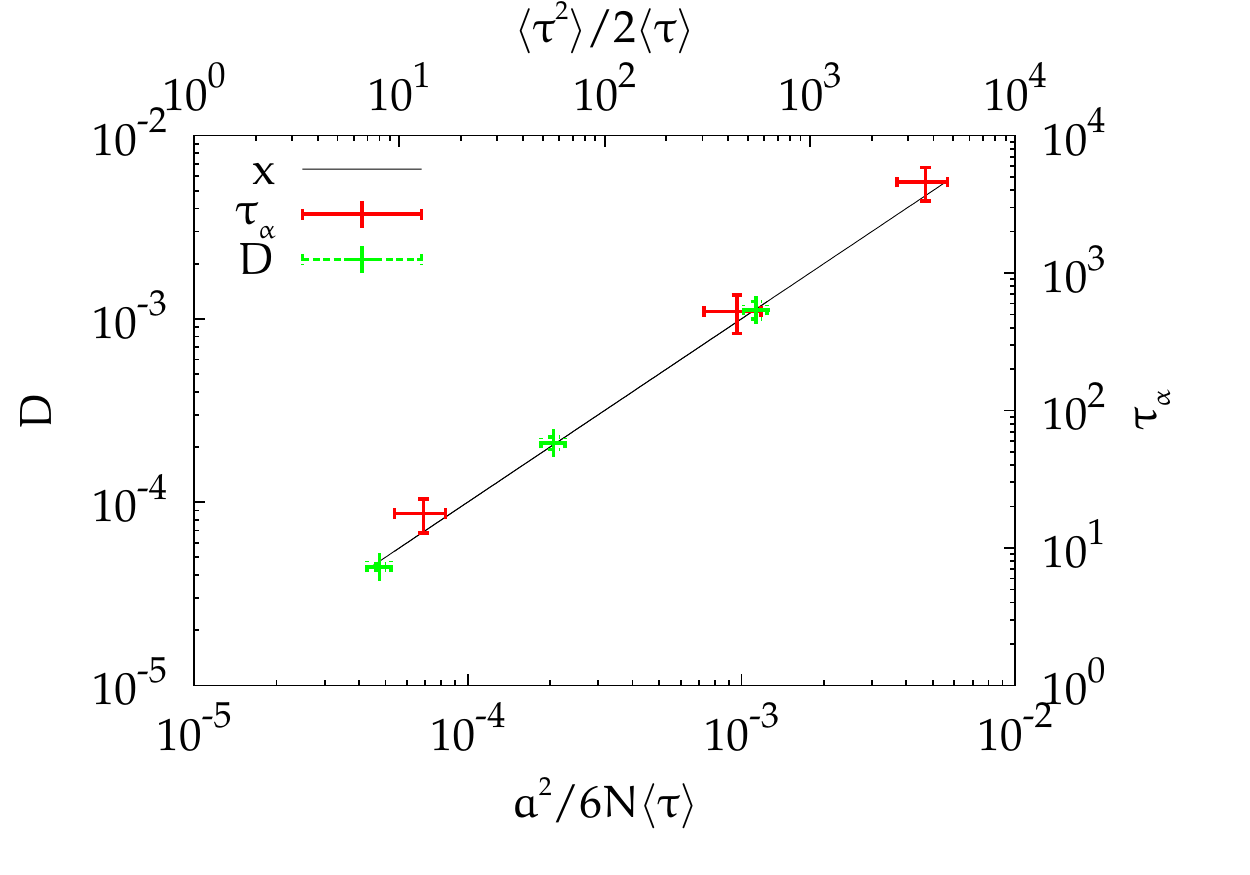}
\caption{Comparison of the absolute values of $D$ and $\tau_\alpha$ from the MB trajectory with
corresponding expressions from
the CTRW PEL description. Included is the diagonal.}
\label{fig: AbsolutBoth}
\end{figure}

As shown in \cite{Rubner2008} the dynamics can be described as a CTRW in configuration space.
Therefore the most relevant transport coefficients like $D$ or $\tau_\alpha$ can be
calculated analytically by solving the Gaussian integrals $\int\Gamma^n(e)p_{eq}(e)\text{d}e$
\cite{Heuer2008a}. Beyond the high temperature limit (no influence of the crossover energy)
results for $\lambda=\kappa=1$ can be found in literature like
the well-known quadratic behavior $\tau_\alpha\propto\exp[\sigma^2(\beta-\beta_0)^2$] with
$\beta_0=-k_\text{entro}$ \cite{Heuer2008,KCMGarrahan2002,Keys2011}. For general $\lambda<1$
the
$T$-dependence of $D$ and $\tau_\alpha$ is not purely quadratic (see
Appendix \ref{app: analytic}). However, for the Stokes-Einstein relation one again obtains a
quadratic dependence
\begin{equation}
\begin{split}
\ln\left[D\tau_\alpha\right]\sim\ln\left[\average{\Gamma^{-1}}
\average{\Gamma}\right] =& \lambda^2\sigma^2\left(\beta+\kappa k_\text{entro}\right)^2.
\end{split}
\end{equation}
In the low temperature limit one gets 
$D\tau_\alpha\sim\tau_\alpha^{\xi}$ with $\xi=2\lambda/(2+\lambda)$ which for the present case
gives $\xi=0.4$.

% A quantitative comparison of the analytical calculations and the MD data is only possible in a
% restricted temperature range. At low temperatures the simulation time for small systems increases
% beyond the computational limit. The lowest temperature studied is $T=0.45$. Below this
% temperature,
% we no longer find accessible states below $e_{cut}\approx \MDEcut$. These states are then
% populated
% rapidly and the DOS deviates from a Gaussian. The former deviations are taken into account during
% the simulations by introducing a power-law cutoff $\exp[-(e-e_{cut})^4]$ below $e_{cut}$. All
% analytical results are valid only for Gaussian distributions. The precise choice of the exponent
% does not seem to be important. At high temperatures, on the other hand, the dynamics are no
% longer activated and depend on the details of the high energy barriers. Furthermore, for $T>0.8$
% $p(e)$ is no longer Gaussian and the mean energy as well as the variance of the DOS does not
% follow
% the Gaussian model prediction. Therefore we can compare the model data in the temperature range
% $T\in[0.45,0.6]$. The above mentioned PEL parameters can be extracted by fitting the former
% functions to the MD rate and DOS (see \cite{Heuer2008} for details). Our results slightly differ
% from literature and can be found in Tab. \ref{tab: parameter}.

In what follows we compare the CTRW/PEL predictions with the actual MD data in the temperature
interval $[0.45,0.6]$. The lowest temperature is close to the mode-coupling temperature $T_c$
\cite{Vogel2004}. More specifically we compare three different types of trajectories:
(i) continuous trajectory from standard MD simulation, (ii) hopping trajectory between the MB
configurations resulting form the same MD simulation, (iii) CTRW trajectory, based on the waiting
times which are generated from the above PEL approach.

In a first step we determine the diffusion constant $D$. By construction (i) and (ii) will result
in identical values of $D$. For the CTRW trajectory one directly obtains $D=a^2/6N\average{\tau}$
where $a^2$ is a weakly temperature dependent value and can be determined independently
\cite{Doliwa2003b}. As shown in Fig. \ref{fig: AbsolutBoth} the temperature dependent first moment
of the waiting time distribution perfectly reflects the $T$-dependence of $D(T)$. Of course,
strictly speaking this comparison is just a consistency check of the CTRW/PEL approach.

\begin{figure}[tb]
\includegraphics[width=\picturewidth]
{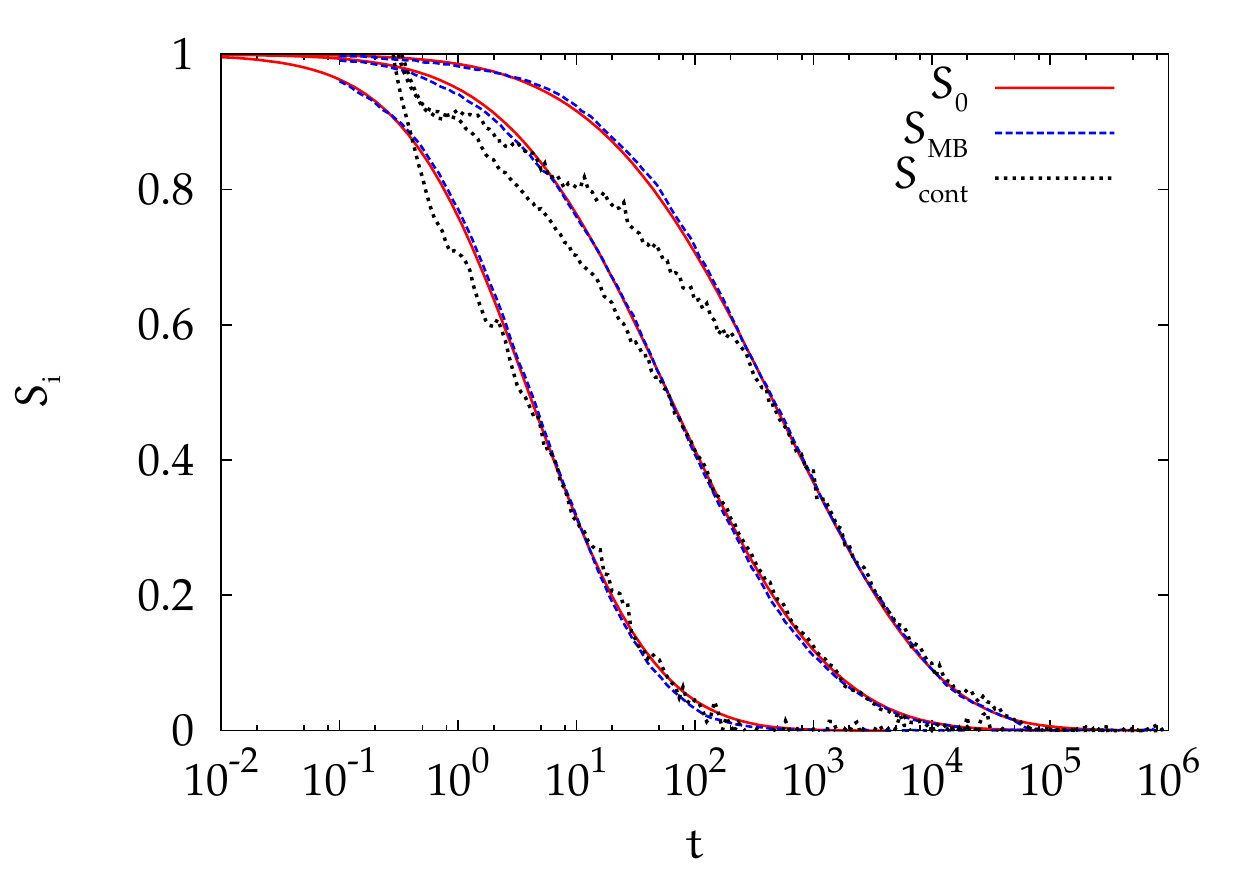}
\caption{Comparison of $S_0(t)$ with $S_{MB}(k=400,t)$ and $S_\text{cont}(k_c=15,t)$ for
$T=0.45,0.5$ and $0.6$.}
\label{fig: vglModellMD-S0}
\end{figure}

\begin{figure}[tb]
\includegraphics[width=\picturewidth]
{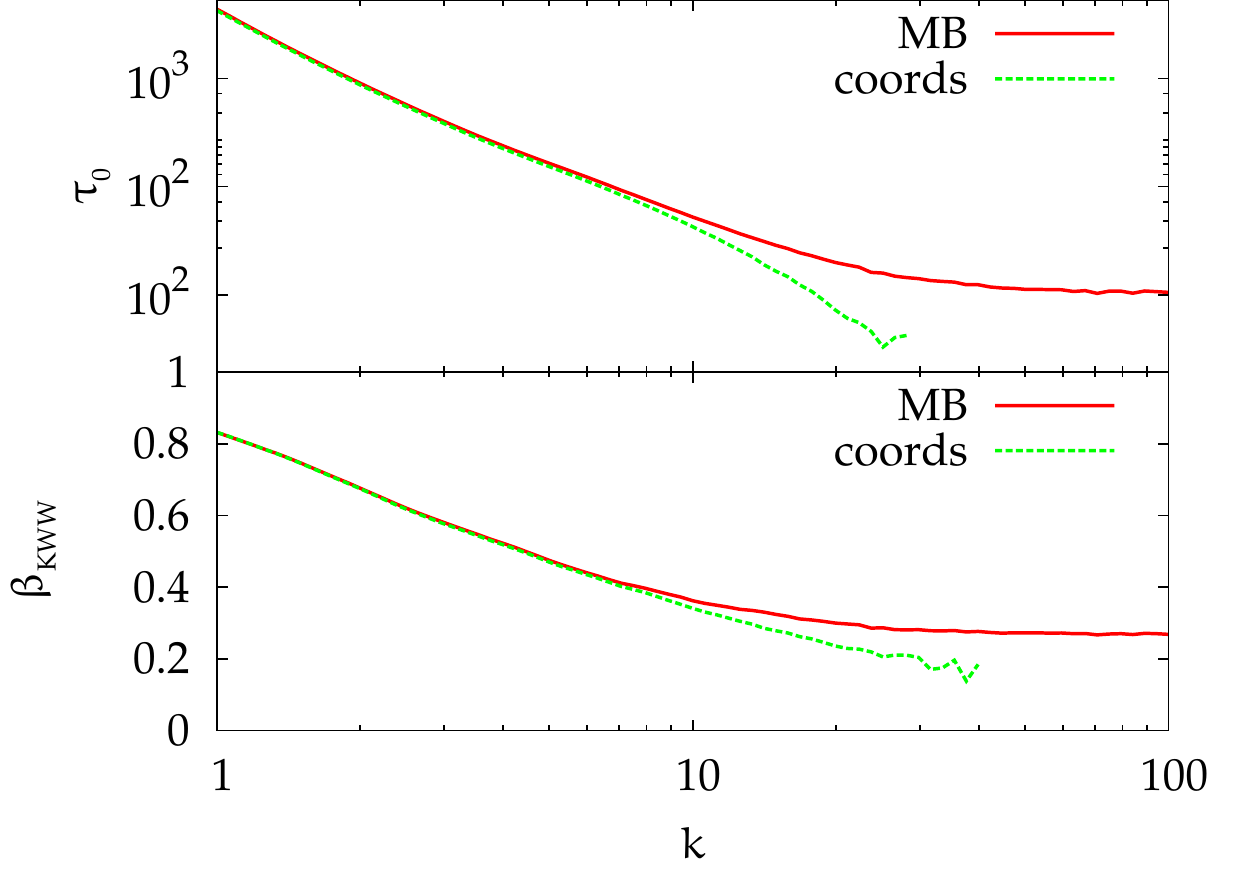}
\caption{$k$ dependence of the fit parameter $\tau_0$ and $\beta_{KWW}$ for the BMLJ65
at $T=0.5$ for MB and real space coordinates.}
\label{fig: BetaKWWCoordsMB}
\end{figure}

In the second step we compare the structural relaxation times $\tau_\alpha$. In the CTRW/PEL
approach $\tau_\alpha$ is conveniently defined via $\tau_\alpha=\int\text{d}t S_0(t)$
where$S_0(t)$ denotes the probability that, starting form a randomly chosen time, the
system  has not performed a relaxation process \cite{Rubner2008}. It can be interpreted as the
persistence time distribution \cite{Berthier2011}. Straightforward calculations yields
$\tau_\alpha=\average{\tau^2}/2\average{\tau}$ \cite{Berthier2005a, Heuer2008}. $S_0(t)$ is shown
in Fig. \ref{fig: vglModellMD-S0} for all considered temperatures. For MD trajectories
$\tau_\alpha$ is typically extracted from the incoherent scattering function $S(k.t)$. For a
comparison of (ii), i.e. the MB hopping trajectory, with (iii) one has to choose a very large
value of the wave vector $k$ so that after one discrete hopping process full decorrelation is
achieved. Choosing $k=400$, one can see in Fig. \ref{fig: vglModellMD-S0} a very good agreement
between $S_{MB}(k=400,t)$ and $S_0(t)$. For $T=0.6$ it turns out that $S_0(t)$ is slightly more
non-exponential. One may speculate that this indicates the onset of anharmonic effects which at
temperature $T=1$ leads to a total breakdown of the PEL approach \cite{Buchner1999,Sastry1998}. For
the determination of $\tau_\alpha$ we fit $S_{MB}$ with a stretched exponential function
$\exp[-(t/\tau_0)^\beta_{KWW}]$. The $\alpha$-relaxation time $\tau_\alpha$ can then be calculated
via $\tau_\alpha=\tau_0(k_c)/\beta_{KWW}(k_c)\Gamma(1/\beta_{KWW}(k_c))$. As expected from the good
agreement in Fig. \ref{fig: vglModellMD-S0} also the $\tau_\alpha$ values of (ii) and (iii) agree
well as shown in Fig. \ref{fig: AbsolutBoth}. More subtle is the quantitative comparison with (i)
because the additional presence of
vibrational and intra-MB processes give rise to additional decorrelation mechanisms for
$S_{cont}(k,t)$. These aspects have to be worked carefully because for large system, see Sect.
\ref{sec: macroscopic}, we are restricted to use the continuous MD trajectories. We start by
fitting $S_\text{cont}(k,t)$ by a stretched exponential for times in the $\alpha$-regime, yielding
fitting parameters $\tau_0^\text{cont}(k)$ and $\beta_{KWW}^\text{cont}(k)$. Their $k$-dependence
for $T=0.5$ is shown in Fig. \ref{fig: BetaKWWCoordsMB}. The same procedure is performed for the MB
trajectory with the corresponding parameters $\tau_0^{MB}(k)$ and
$\beta_{KWW}^{MB}(k)$. $\tau_{0}^{MB}(k)$ and $\tau_{0}^{coord}(k)$ (calculated from MB
respectively real space coordinates) correspond to each other up to a length scale of $k\approx10$.
For larger values
of $k$ one finds vibrational dynamics and $\tau_{0}^{MB}(k)$ saturates, while
$\tau_{0}^{coord}(k)$
further decreases. The plateau values of  $\tau_{0}^{MB}(k)$ and$ \beta_{KWW}^{MB}(k)$ we are
interested in can now be estimated by finding the value of $k_c$ where $\tau_{0}^{coord}(k_c)$ and
$\beta_{KWW}^{coord}(k_c)$ matches with the MB plateau. We found $k_c\approx\kc$.
It is promising and non-trivial that for the same value of $k$ both parameters can be recovered. 
$k_{c}$ is slightly temperature dependent, e.g. for $T=0.6$ one gets $k_{c}\approx 14\pm2$. As
shown in Fig. \ref{fig: vglModellMD-S0} $S_\text{cont}(k=k_c,t)$ agrees indeed very well with
$S_{MB}(k=400,t)$. Note that for this comparison the short time decay of $S_\text{cont}(k=k_c,t)$
has been scaled out. An equivalent scaling has been already used in Ref. \cite{Schroder2000} where
continuous and IS trajectories had been compared.

\section{The macroscopic system}\label{sec: macroscopic}

\subsection{General}
With increasing number of particles the properties of MB becomes less useful due to the following
reasons:
1) For the minimum system consecutive MB transitions are uncorrelated, because that during a
transition the mobile particles are more or less equally distributed over the sample. In large
systems,
due to the dynamic heterogeneity, transitions are typically performed by the same mobile particles.
Thus, the real space information about the location of a relaxation process is getting important.
2) Due to the many rearranging regions in the sample, the entire system performs a transition in any
given fixed time interval, leading to a $\delta$-peaked waiting time distribution.
3) $p_{eq}(e)$ is narrowing when increasing the system size. As a consequence the relation between
$\Gamma$ and $e$ is smeared out and the escape rate is much less correlated with the energy.
% 4) The total distribution of rates becomes small compared to the distribution of rates for a
% single
% energy (for comparison: At $T=0.5$ the width of $p(\log\Gamma)$ is $\sigma(N=65)\approx 4.7$,
% 	$\sigma(N=130)\approx 3.5$ and $\sigma(N=260)\approx 2.4$, giving rise to the assumption
% that for $N$ larger than $\approx 600$ the relation is inverted.)

As reported in \cite{Rehwald2010}, local waiting times remain a precise measure for the dynamics
even in large systems. This means that CTRW results can also be applied in large systems. For the
comparison with the $\alpha$-relaxation time $\tau_\alpha$ of the large system we have used
$S_\text{cont}(k_c,t)$ since $S_{MB}(k,t)$ can no longer be defined (see discussion above).

\subsection{Evidence for coupling in a non-equilibrium configuration}
  
So far nothing is known about the size or shape of subsystems, furthermore the kind of coupling
is unclear. But how can one identify coupling processes? When studying large systems one faces the
problem of strong "dynamic noise" which makes it difficult to distinguish between different coupling
events. To minimize the dynamic noise, we prepared a very immobile configuration of a 520 particle
system by copying a very stable $N=65$ structure at $T=0.5$. From this non-equilibrium
structure we calculate the iso-configurational ensemble (IC) to study the distance dependence of
possible coupling events between adjacent regions.

\begin{figure}[tb]
\includegraphics[width=\picturewidth]
{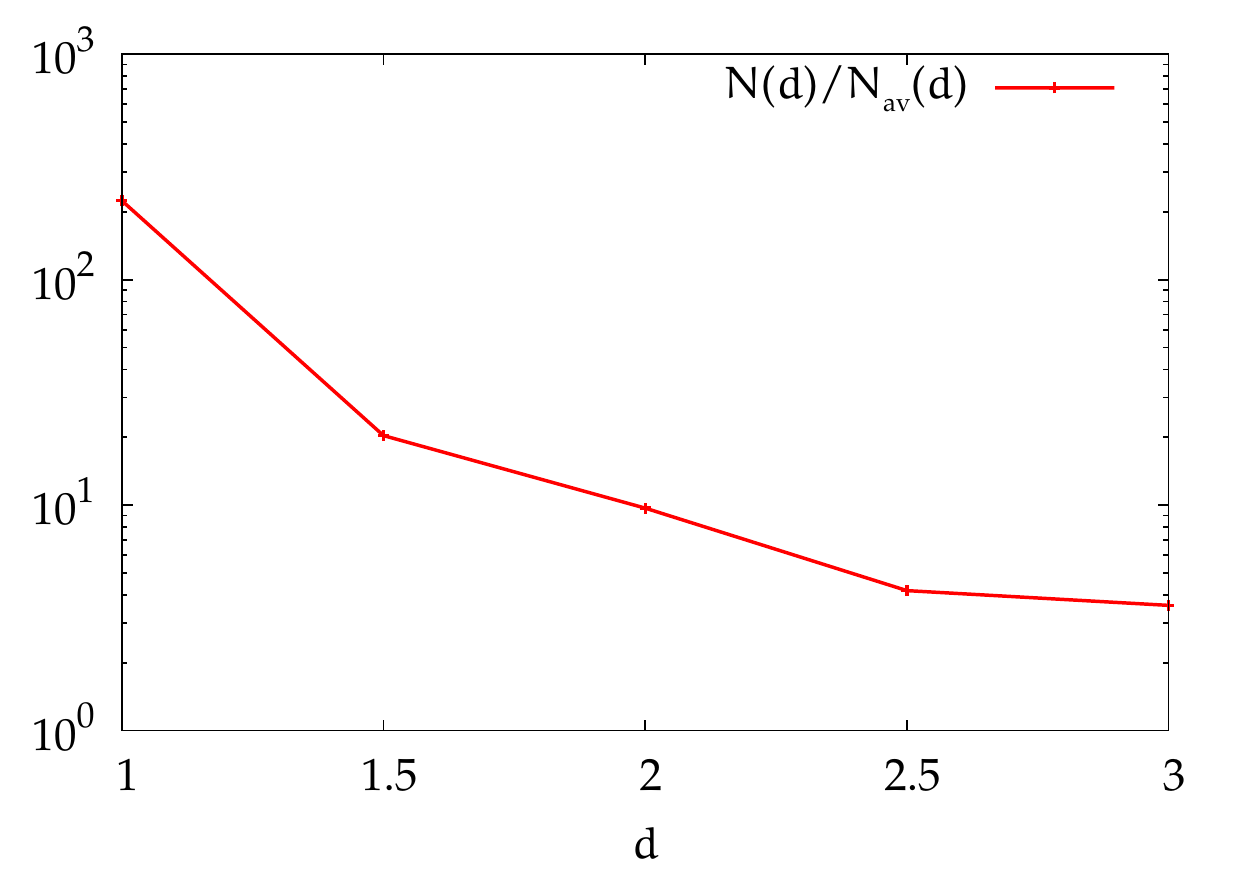}
\caption{Number of first events in the distance $d$ normalized by the average number of particles
one would expect from the radial distribution function.}
\label{fig: MSD non-equi N}
\end{figure}

At the beginning of each simulation two distinct behaviors can be identified: The $a$-particles are
organized in a stable $a$-matrix which allows string like motion of $b$ particles without changing
the structure significantly. Hence we neglect $b$ particles in the local event calculation. If an
$a$-particle changes its position, it is either a subset of particles exchanging their positions
inside the unaltered matrix or it reflects a significant local change of the matrix structure. The
first process can be detected via local events but is irrelevant for the decay of the initial
structure. The latter gives rise to define \emph{structural} events: 
When an $a$-particle performs a local event at time $t$, we calculate the time averaged coordinates
of the tagged particle and the first shell particles at $t\pm\Delta t$ with $\Delta
t\approx\tau_\alpha$. To determine if the structure of the first shell changes significantly during
the exchange process, we calculate the squared displacement $MSD=\sum_i (r_i(t+\Delta
t)-r_{i'}(t-\Delta t))^2$ of particle positions before and after the central event. The index $i'$
corresponds to the particle residing closest to the initial position of particle $i$
(allowing permutation). In equilibrium one obtains $\langle MSD \rangle\approx 6$, the distribution
is
very similar to a Gaussian with a variance close to 2. For the definition of a structural event we
use a threshold of 2, above which we call the event structural. The precise choice of the threshold
does not change the result.

Before we present the MD results we first discuss possible effects in a small model system: Consider
three independent systems with rates $\Gamma_i=\Gamma$. After the first
(the left) system changes its state as a consequence of a relaxation process we now discuss the
location of the \emph{next} relaxation process. Here we concentrate on the middle and right system.
More specifically we discuss the ratio $\average{N_m}/\average{N_r}$ ($\average{N_i}$: the number
of next relaxation processes in system $i$) as well as
$\average{\tau_{1,2}^{(m)}}/\average{\tau_{1,2}^{(r)}}$ ($\average{\tau_{1,2}^{(i)}}$ denoting the
average waiting time between the initial process of the left system and the next relaxation process
in system $i$).
Since both systems have the same rate one has
$\average{N_m}=\average{N_r}=1$ and $\average{\tau_{1,2}^{m,r}}=1/2\Gamma$, yielding
$\average{\tau_{1,2}^{(m)}}/\average{\tau_{1,2}^{(r)}}=1$. Now we introduce a coupling mechanism
where
the initial relaxation enables the central system to acquire the rate $\Gamma_m=2\Gamma$ whereas the
rate
$\Gamma_r$ of right system remains unchanged. Then one naturally has
$\average{N_m}/\average{N_r}=2$ and
$\average{\tau_{1,2}^{(m,r)}}=1/3\Gamma$ for both systems so that again
$\average{\tau_{1,2}^{(m)}}/\average{\tau_{1,2}^{(r)}}=1$.
The situation changes if one assumes a distribution of rates $p(\Gamma)$, e.g. $\Gamma_m=2\Gamma$
and $\Gamma_m=4\Gamma$ both with probability 0.5. In this case on has
$\average{N_m}/\average{N_r}=3$ and
$\average{\tau_{1,2}^{(m)}}/\average{\tau_{1,2}^{(r)}}=11/12<1$. The ratio decreases with
increasing width of $p(\Gamma)$.

For the BMLJ system this argument implies that as a consequence of a local coupling mechanism it
is by far more likely that the second relaxation process occurs close to the initial one. Averaging
over the IC ensemble would observe a significant distance $d$-dependence $N(d)$. The corresponding
distance dependence of  $\average{\tau_{1,2}}$ gives information about the strength of the
rate fluctuations. 

In Fig. \ref{fig: MSD non-equi N} we show $N(d)$ normalized by the average number of
particles $N_\text{av}(d)$ one would expect from the radial distribution function $g(r)$. The curve
decays
monotonically until a plateau value is reached for $d>2.5$. As discussed for the model system this
is a clear evidence for coupling processes. It would contradict the statistical case where no rate
fluctuations are present. The curve roughly reaches a plateau value for $d>2.5$ giving rise to the
assumption that only particles within a sphere of radius $r\approx2.5$, which is a little bit
larger than the minimum system, are directly affected by the first relaxation event. Beyond this
sphere, particles hardly recognize the first event, at first, and the changes of the rate become
negligible. The discussion of $\average{\tau_{1,2}}(d)$ can be found in Appendix \ref{sec:
tau12}.

The main contribution of the coupling is therefore between adjacent minimum systems so the coupling
range can be restricted to such systems. These results reflect the underlying coupling processes and
do not depend on the details of the definition of relaxation events.

  \begin{figure}[bt]
  \includegraphics[width=0.8\picturewidth]
  {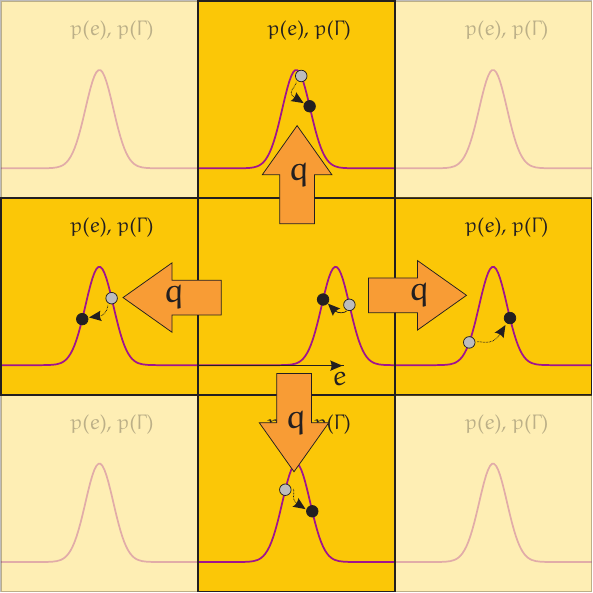}
  \caption{Schematic sketch of the physical scenario: When the central system relaxes independently
  (active process, black arrow) the adjacent system may change its mobility without changing its
  state (passive process, dashed arrow). The role of $q$ is discussed in the text.}
  \label{fig: PhysicalScenario}
  \end{figure}

\subsection{Physical picture}
  
  To illustrate our physical picture of supercooled liquids we make use of the equilibrium
distribution of rates $p(\Gamma)$ introduced in \cite{Rehwald2010a}. Since thermodynamic properties 
\cite{Doliwa2003} as well as the diffusivity \cite{Rehwald2010}
do not change upon increasing the system size (for $N>N_\text{min}$) we strongly
suggest that the local equilibrium distribution $p(\Gamma)$ of the mobility does not change either.
Therefore we assume that at any arbitrary time a macroscopic system can be decomposed into
microscopic subregions of the size of the minimum system which can be described by the properties of
their PEL, i.e. by values of $e$ and $\Gamma$.
When a subregion relaxes, this is called an active process. 
If one describes a real system by a decomposition into elementary systems, one has to include
interactions between these subregions, which is
what we call coupling.
These interactions allow the subregions to change the rate while keeping
their overall Boltzmann distribution $p(\Gamma)$ (passive process), see Fig. \ref{fig:
PhysicalScenario}.
Now consider an immobile region adjoining to a rearranging mobile region: The adjacent relaxation
enables rate fluctuations in the slow sample, on average leading to a higher rate. This scenario
directly explains the lack of slow regions in large model glass former compared to the minimum
system \cite{Rehwald2010}.
The fluctuations themselves are interpreted as a coupling mechanism which leads to facilitation
like dynamics, sometimes viewed as a hierarchical process \cite{Keys2011}.
The precise realization of
passive processes will be discussed in the next sections. This kind of coarse graining and coupling
is in principle captured in the kinetically constrained models, but two important differences
remain: 1) All subsystems can always relax independently and 2) a subsystem already contains the
complete macroscopic thermodynamics as also realized in the mosaic approach.

\subsection{The coupled landscape Model}
  
Since we have shown that the minimum system can be described by PEL parameters and the CTRW
formalism and that localized coupling processes play a major role in supercooled liquids, we
bring together both ingredients: We interpret a macroscopic glass former as a set of weakly
coupled elementary systems which can be described by their PEL. The elementary systems (ES) are
arranged
on a square lattice, and their time evolution can be simulated as in \cite{Heuer2005}. After each
active process all coupled adjacent systems are allowed to perform a passive process. We denote this
approach as \emph{coupled landscape model} (CLM). It is somehow a minimal model
based on the PEL. It has to fulfill the condition that the thermodynamic
properties as well as the diffusion coefficient $D$ remain unchanged when increasing the system
size.

Dynamic coupling enables so called \emph{passive} processes: Due to active processes of adjacent
regions, the mobility (the rate) of an ES can also change without performing a relaxation process. 
The presence of passive processes must not change the equilibrium distribution $p(e)$. Therefore
the transition probability  $\pi(e_{old}\rightarrow e_{new})$ to move from state $e_{old}$ to state
$e_{new}$ for a passive process has to fulfill
\begin{equation}
\int p(e_{old})\pi(e_{old}\rightarrow e_{new})\text{d}e_{old} = p(e_{new})\quad .
\label{eq: dynamic coupling}
\end{equation}
The latter condition is needed for the probability interpretation. In this paper we
focus on the most simple case $\pi(e_{old}\rightarrow e_{new})=p(e_{new})$, what we call Boltzmann
coupling:
With
the probability $q$ the new rate is simply chosen from the equilibrium distribution. Another simple
case is the Gaussian coupling where $e_{new}$ and $e_{old}$ are correlated, meaning that only small
energy changes are allowed. With mean $\mu$ and variance $\sigma$ of the DOS one can define
$q(e_{old},e_{new}) =1/\sqrt{2\pi s^2}\exp{\left[-(e_{new}-ae_{old}-b)^2/2s^2\right]}$ with
constants $a=\sqrt{1-s^2/\sigma^2}$ and $b=\mu(1-a)$.

\section{Results}\label{sec: results}
In this section we fix the coupling constant $q$ and show that this model
allows a non-trivial reproduction of the most important transport coefficients. This comparison is
based on observables, which are well-defined in the MD simulation and the CLM.
Here we estimate the coupling strength $q$ based on the $\alpha$-relaxation time $\tau_\alpha$.

In previous work this kind of dynamical coupling was first estimated within a mean-field approach
from the reduction of $\tau_\alpha$ when going from $N=65$ to $N=130$ particles by a linear
expansion in $q$ of the relaxation time $\tau_\alpha(N)$ of the entire system \cite{Rehwald2010}.
Here, we follow a more general route which, first, is not based on linear expansion and, second,
reflects the transition to macroscopic systems. For the simulations of the CLM we have used a $3^3$
system with the parameters listed in table \ref{tab: parameter}.

\ifpdf
\subsection{\texorpdfstring{$\alpha$}{alpha}-Relaxation}
\else
\subsection{$\alpha$-Relaxation}
\fi

\begin{figure}[tb]
\includegraphics[width=\picturewidth]
{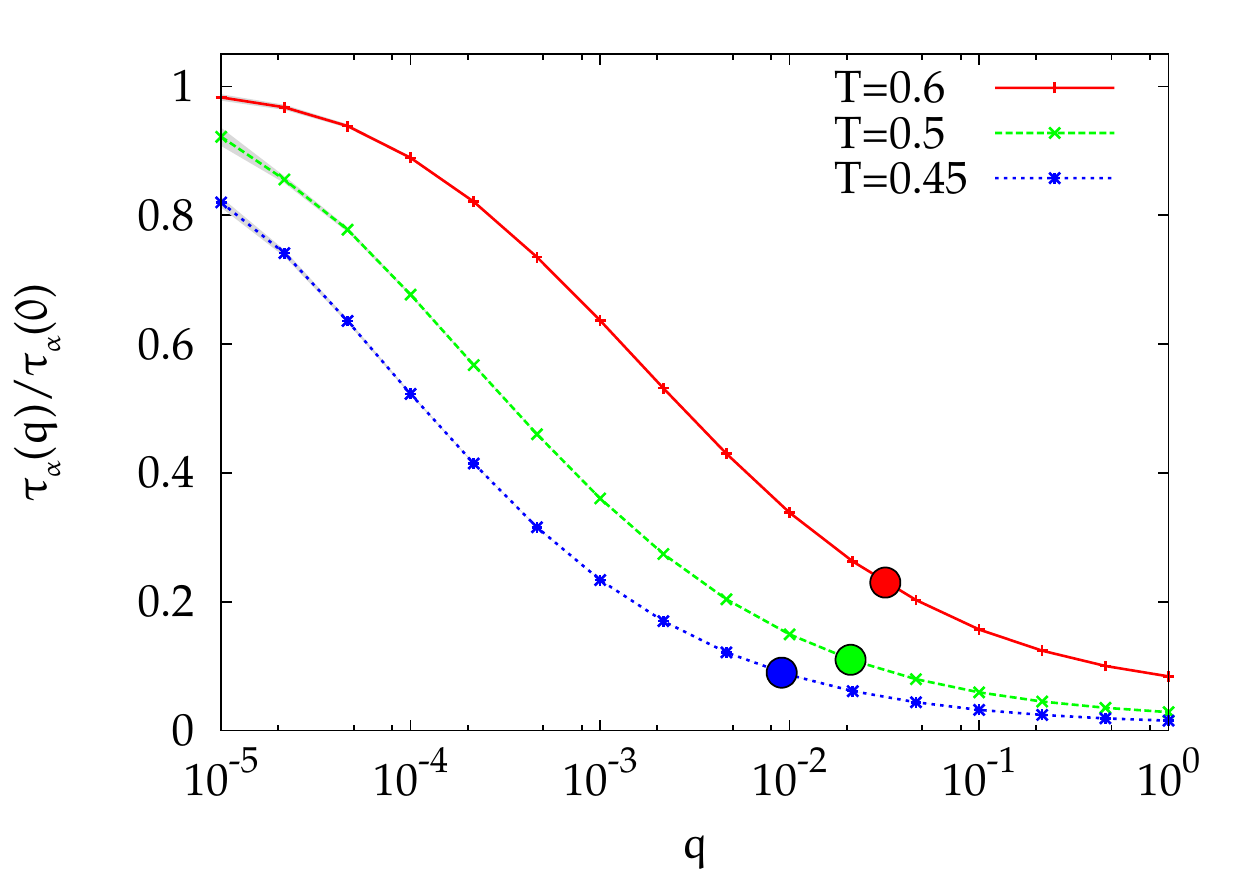}
\caption{Relative reduction of $\tau_\alpha$ vs. coupling probability $q$ for different
temperatures. The solid dots correspond to MD results
($\tau_\alpha(8320)/\tau_\alpha(65)=0.32,0.11,0.08$ for $T=0.6,0.5,0.45$), the gray region
corresponds to the error intervals.}
\label{fig: TauAlphaQ}
\end{figure}

Again we will use finite size effects of $\tau_\alpha$ to estimate $q$. We increase the system size
from $N_\text{min}$ to $N=8320$ and calculate $\tau_\alpha(N)$ from the MD data. $N=8320$
corresponds to the macroscopic limit for these temperatures. In the CLM one can
model
this scenario by adding additional ESs. If an ES corresponds to the smallest system in the MD, then
one can directly compare the relative reduction of $\tau_\alpha$ from an equilibrium simulation. For
the dynamical interaction all systems sharing a boundary with the active can experience a passive
process.

The MD results show minor finite size effects for $D$ \cite{Rehwald2010}. In several papers this
phenomenon is related
to hydrodynamic interactions of the sample with its images due to the periodic boundary conditions
\cite{Heyes2007,Heyes2007a}. However, these hydrodynamic interactions are not captured by the model
($D$ remains constant under coupling). To compare MD and model data, we first have to remove the
hydrodynamic effects by scaling $D_N$ and $\tau_\alpha$ by $D_N/D_\text{min}$. The correction of
$\tau_\alpha$ for the largest systems studied is about one order of magnitude smaller than the
observed finite size effect itself, so these hydrodynamic effects are negligible for this study.

We determine $q$ by the condition that the reduction of $\tau_\alpha$ of the BMLJ system
and the CLM exactly match. In Fig. \ref{fig: TauAlphaQ} we show the
results for three different temperatures. From this analysis we get a temperature dependent
coupling constant $q(T)$. In the chosen temperature interval, $q$ decreases by a factor of roughly
3. For small and intermediate values of $q_\alpha$ the empirical formula
$\tau_\alpha(q)=\tau_\alpha(0)/(1+c\sqrt{q})$ fits the data very well. We mention by passing that
the reduction does not seem to be analytical in the $q\rightarrow 0$ limit, unlike the linear
dependence in $q$ we used earlier in a mean-field approach \cite{Rehwald2010}.

\begin{figure}[tb]
\includegraphics[width=\picturewidth]
{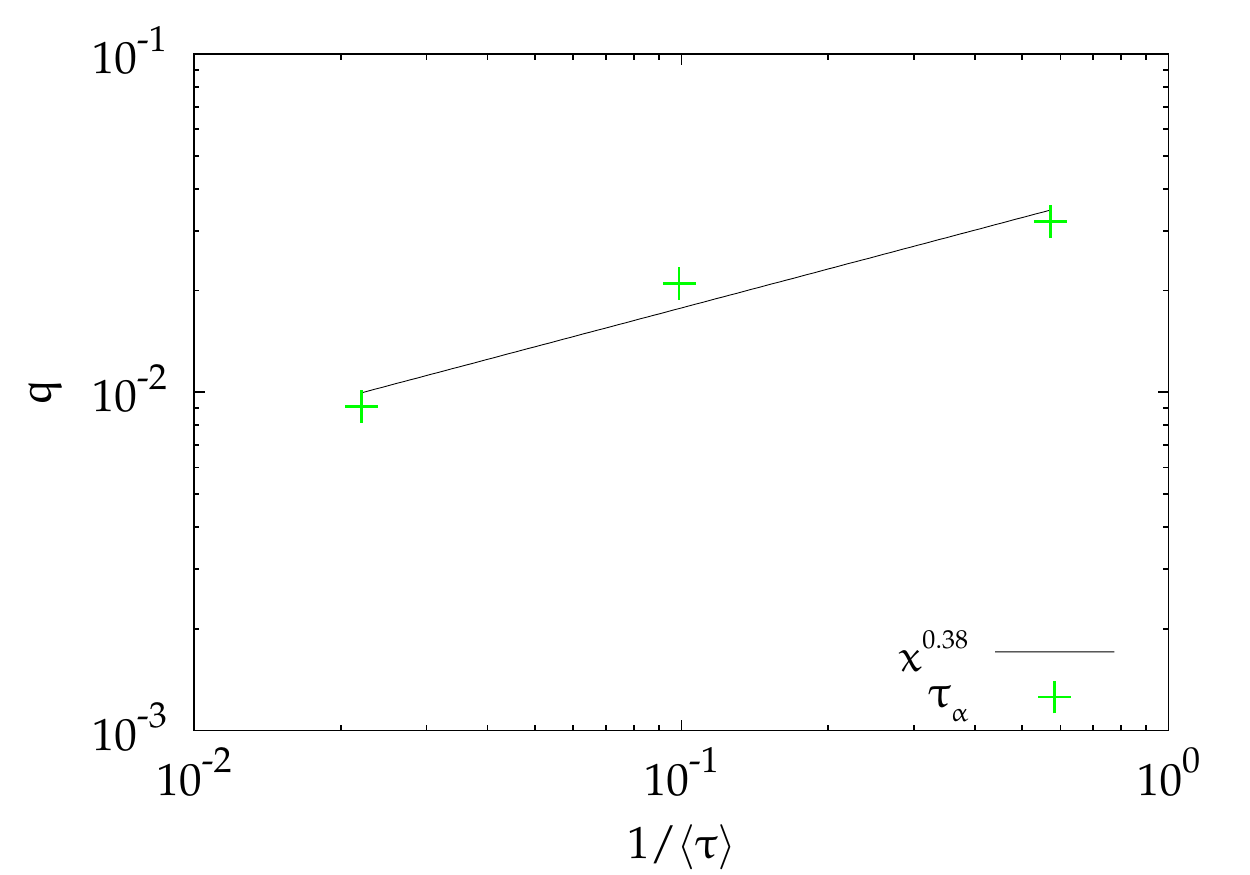}
\caption{Coupling constant $q$ vs. average waiting time for different temperatures. The black solid
line is a power law as guidance to the eye.}
\label{fig: QofTau}
\end{figure}

The estimation of $\tau_\alpha$ provides data for the temperature dependence of $q$. To answer the
question how the significance of the facilitation procedure evolves with temperature we compared $q$
with typical timescales in the system. In Fig. \ref{fig: QofTau} a sublinear scaling with
$1/\average{\tau}$ can be found, meaning, that with decreasing temperature the number of
successful passive processes decreases. At the same time the heterogeneity of the elementary
systems increases making a passive process more effective. 
Both mechanisms compete and we can not determine the total effect on the significance of
facilitation.
A similar behavior was also found in granular systems
\cite{Candelier2010a}, where the number of facilitation processes decreases when increasing the
density up to the granular glass transition.

\begin{figure}[b]
\includegraphics[width=\picturewidth]
{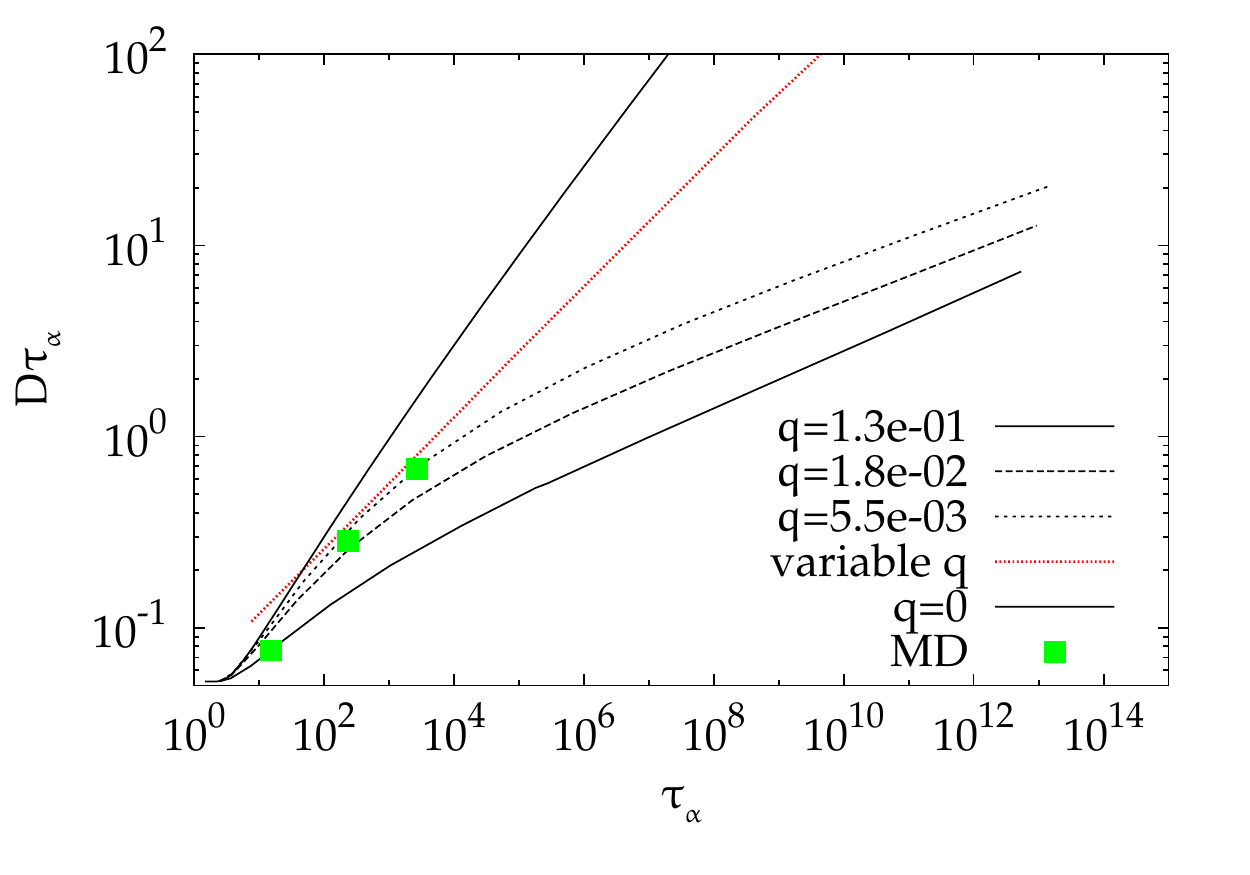}
\caption{Effect of the passive processes on the Stokes-Einstein relation: Black lines are calculated
with fixed coupling constants, the red one corresponds to a temperature dependent coupling
constant, $\propto \average{\Gamma}^{1/2}$,
the green points correspond to the MD data.}
\label{fig: SERallQInclMDData}
\end{figure}

Since $\tau_\alpha$ depends on the first two moments of $\varphi$ and the changes of 
$\average{\tau}$ due to the coupling mechanism are negligible by construction, the reduction of
$\tau_\alpha$ in 
Fig. \ref{fig: TauAlphaQ} mainly reflects the strong decrease of $\average{\tau^2}$. In Fig.
\ref{fig: SERallQInclMDData} we show the violation of the Stokes-Einstein relation to
demonstrate the influence of the coupling processes on the dynamics at different temperature. The
upper black solid line is the analytical result without coupling. The
dashed
curves correspond to CLM data with constant coupling strength for different values of $q$ and the
red line to  CLM data with a temperature dependent $q$  with an exponent of $\xi=0.5$. The
decoupling of $D$ and $\tau_\alpha$ is obviously strongly reduced by the coupling and $\xi$ does not
depend on $q$
in a first order approximation. For low temperatures it reaches the value for large $q$ of
approximately $0.175$. The prefactor shows a strong $q$ dependence. If one assumes a
temperature dependent $q$, for example $q\sim\average{\Gamma}^{1/2}$, see below for the motivation
for this choice, it is also possible to reach
intermediate values for $\xi$. In literature one finds values between $0.05<\xi<1/3$, also values
around $0.5$ are reported for polymers \cite{Mallamace2010, Andraca2008, Voronel1998}).

\begin{figure}[tb]
\includegraphics[width=\picturewidth]
{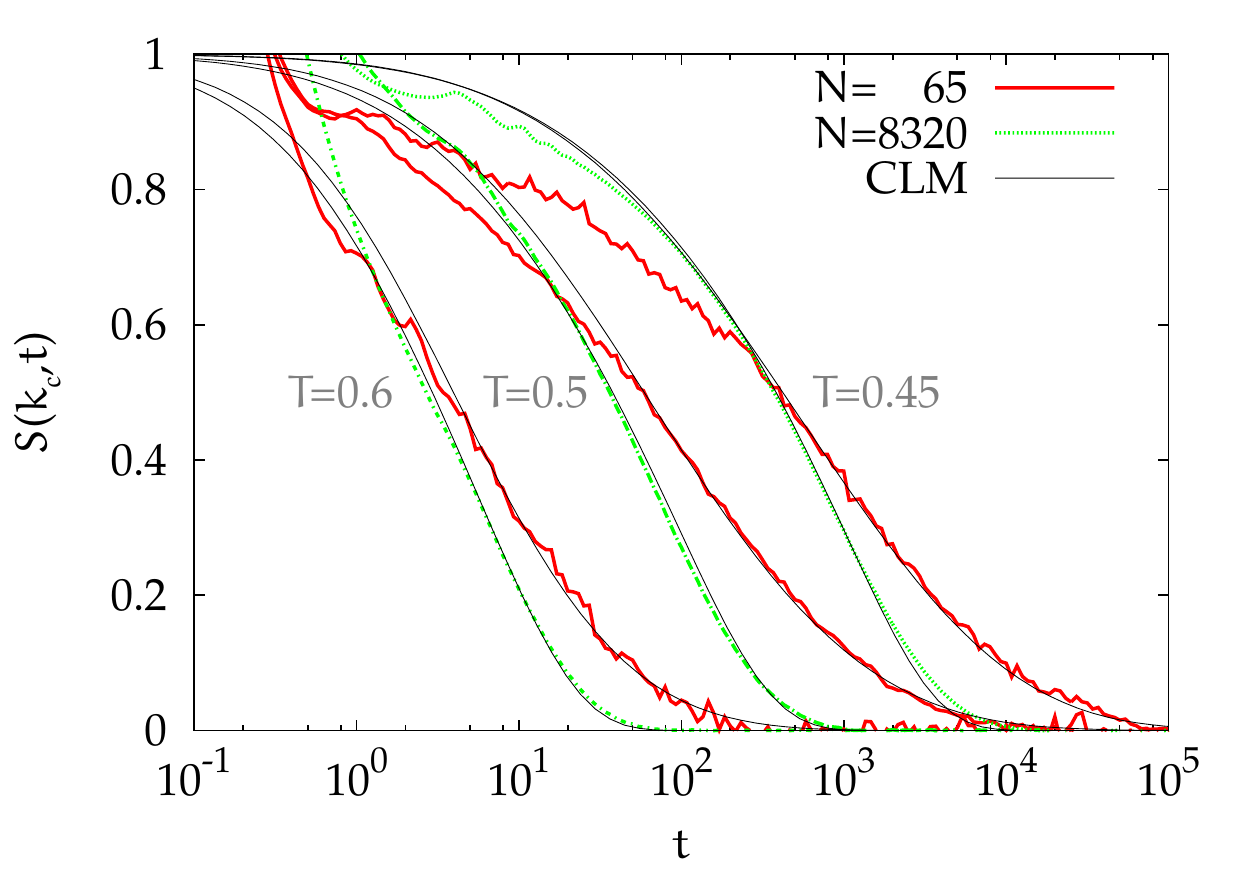}
\caption{Overview over $S(k_c,t)$, $S_\text{cont}(k_c,t)$ (both scaled for optimal overlap) and
$S_0(t)$ from the CLM for $T=0.6,0.5,0.45$ and $N=65,8320$.}
\label{fig: vglModellMD-S0 8320}
\end{figure}

In Fig. \ref{fig: vglModellMD-S0 8320} one can see that the CLM results, $\tau_\alpha$ and also the
non-exponentiality parameter $\beta_{KWW}$, match very well with the BMLJ8230 at $k_c$.  At long
times the relaxation of the CLM is a little bit faster
than in the MD, at shorter times we find the inverse relation. However, the mismatch is small and
can be rationalized by possible heterogeneity of the coupling: For the simulation we have used a
fixed coupling constant. One can imagine, that in the microscopic system one has a distribution of
coupling constants. The
presence of smaller $q$ values immediately leads to slower relaxation at long times, while the
presence of larger coupling constants will increase the decay of the relaxation function at short
and intermediate times.

\subsection{Discussion}

\begin{figure}[tb]
\includegraphics[width=\picturewidth]
{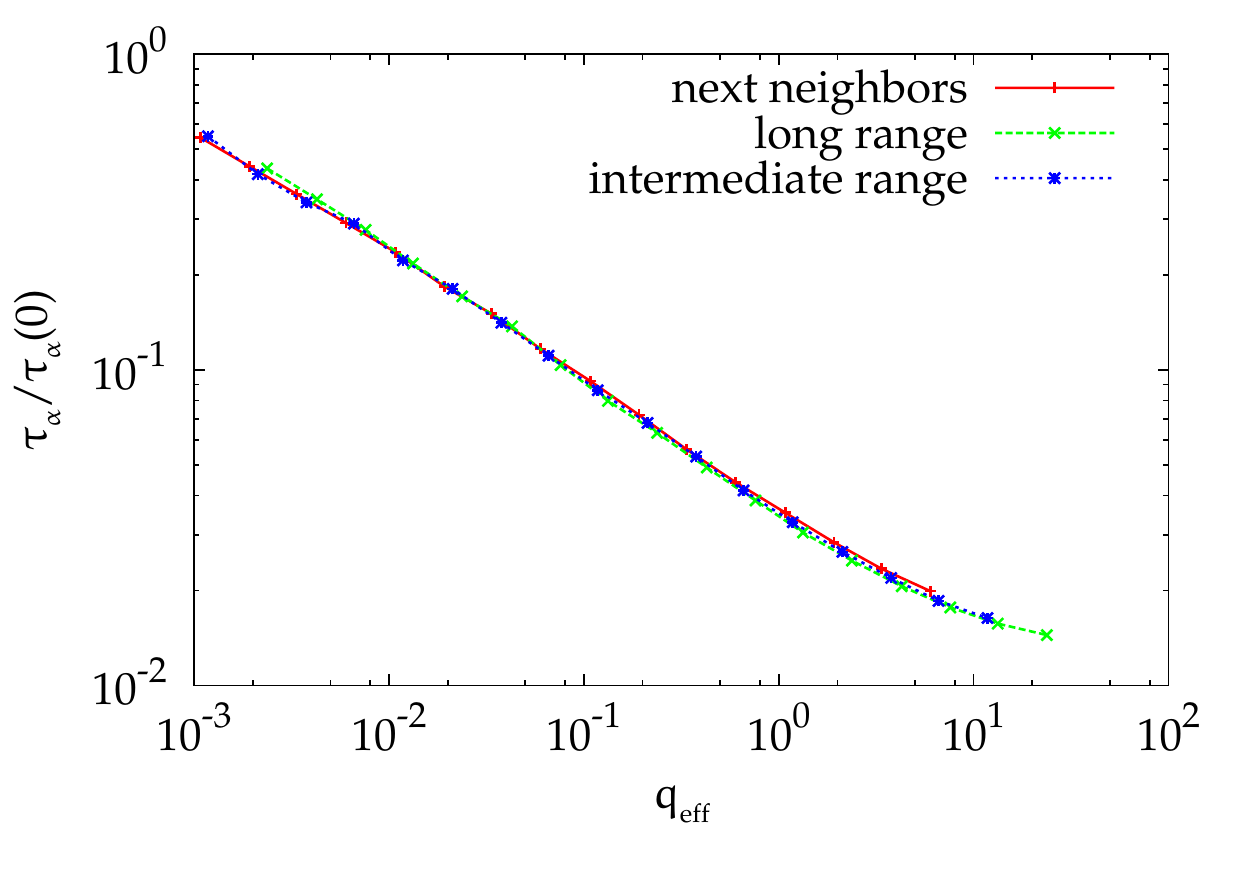}
\caption{Relative reduction of $\tau_\alpha$ vs. coupling probability $q_{\mathrm{eff}}$ for
different neighbor geometries. Long range corresponds to $r_\text{max}=5$, intermediate range to
$r_\text{max}=2$.}
\label{fig: TauAlphaElastic}
\end{figure}

First we want to check how different choices for the geometry of possible passive processes
influence our results. Let $N_\mathrm{eff}$ denote the number of affected systems. In the last
section an active process triggers all neighboring ES (sharing a boundary), i.e.
$N_\mathrm{eff}=2d$. Other realizations are possible, e.g. the elastic case: In
\cite{Bedorf2010} the authors studied finite size effects of the mechanical loss in thin films of a
metallic glass and found that the loss vanishes below a certain thickness of the film. In their view
the Eshelby stress field around a plastic zone is up to three times larger than the excitation
itself. In dielectric relaxation experiments on glycerol it was found that at low temperatures the
relaxation becomes nonlocal and $\tau_\alpha$ decreases with system size \cite{Pronin2011}. The
latter effect was not observed in the CLM. However, these effects give rise to some long-range
elastic interaction between excitations: An ES can be facilitated with the
probability $p(r)\sim r^{-n}$ for $r<r_\mathrm{max}$. Hence, every system
contributes with $r^{-n}$ to the number of effective interacting systems $N_\mathrm{eff}$. This
value can be calculated as $N_\mathrm{eff}=\sum_i r^{-n}$. As the coupling constant $q$
only describes  the probability to perform a passive process for one system, it is useful to
introduce the effective coupling strength
$q_\mathrm{eff}=N_\mathrm{eff}q$. Fig. \ref{fig: TauAlphaElastic} shows $\tau_\alpha$ vs.
$q_\mathrm{eff}$ for different interaction lengths $r_\mathrm{max}$ and illustrates the importance
of $q_\mathrm{eff}$ as an effective parameter. All curves almost lie on a master curve,  so that
the distance dependence of the coupling effects only play a very minor role.

One of the most important results is the temperature dependence of $q$. How can this be
interpreted on the level of a single MB transition? One can think of the following physical
scenario: Every state has an intrinsic resistance against being facilitated leading to an
energy-dependent coupling constant $q(e)$: the lower the energy, the more stable is the MB. This
additional factor has, of course, to be considered
when choosing the new state to obtain detailed balance. A general $e$ dependence cannot be handled
analytically, but for $q(e)=\tilde{q}\Gamma(e)$ one can use the transition probability
$\pi\propto \Gamma\varphi$ which generates the correct statistics. If we now calculate $q$ for
different temperatures so that
$\tau_\alpha(q)/\tau_\alpha(0)$ exactly matches with $\tau_\alpha(\tilde{q})/\tau_\alpha(0)$ we
find some temperature dependence for $q$.
Effects along this line may thus
rationalize the observed temperature dependence.

\subsection{Conclusion}
In a first step we have analyzed in detail how the dynamics of a small BMLJ system with 65 particles
can be expressed in terms of PEL parameters. A key step in this endeavor is the fragmentation of the
configuration space into MBs, each of which is characterized by an energy. Furthermore, in a simple
PEL
approach the rate to escape from a given MB is completely characterized by its energy and,
more
specifically, can be expressed by an effective barrier height and an entropic prefactor. Then it is
possible to predict the dynamics for all temperatures. For this present case one additional aspect
has to be taken into account: a given energy does not fully characterize the escape rate but
the escape rate rather follows a relatively narrow distribution as expressed by a log-normal
distribution.
Intuitively, this reflects the fact that the BMLJ65 system can be interpreted as a superposition of
approx. two elementary subsystems \cite{Heuer2005}. All PEL properties can solely be inferred from
an appropriate analysis of the simulation data.

As a consistency check one can estimate, e.g., the diffusivity or the structural relaxation time.
Indeed, an excellent agreement is found. Stated differently: the dynamics of the BMLJ65 system is
very well understood in terms of the PEL. Of course, due to the discretization the PEL approach
cannot resolve properties of the $\beta$-relaxation.

In principle one can perform the same MB discretization also for much larger systems. However, in
this case the total energy looses the tremendous information it has for small systems. Because a
large system can be decomposed into (roughly) independent smaller systems the same total energy may
result from many different realizations. As a consequence many different relaxation rates to escape
from this MB are possible. Whereas for small systems the distribution of rates for a given energy is
very small as compared to the total distribution of rates (as inferred from the different waiting
times) this relation is inverted for large systems. Furthermore, for large system the mapping on the
CTRW description is invalidated because of spatial relations.

Therefore one has to complement the PEL approach of small systems by a new
concept which takes into account the fact that different regions of the glass-forming system act
somehow independently. Our CLM approach is guided by the observation of causal relations between
different relaxation processes. Starting from an immobile non-equilibrium system these causal
relations can indeed been identified. Via Monte Carlo simulations of the collection of BMLJ65
systems, interacting via the appropriately chosen coupling rules, relevant observables can be
determined and compared with the properties of the large BMLJ systems. Interestingly, the precise
choice of the coupling rule is not important for the properties of $D$ and $\tau_\alpha$. We
carefully identified the wave vector for which the $\tau_\alpha$-value has to be extracted in order
to be compatible with the generic structural relaxation time from the CTRW approach. Comparison of
the CLM with the actual system allows one to identify a coupling constant $q$ which turns out to be
temperature dependent. One may envisage that the resistance of a local region to change its
relaxation rate as a consequence of close-by rearrangements becomes larger for lower local energies,
then this $T$-dependence follows quite naturally. Most importantly, the variation of the whole shape
of the incoherent scattering function $S_\text{cont}(k,t)$ when going from small to large systems
can be
reproduced by the CLM approach, i.e. by adjusting a single parameter. As a consequence the PEL
parameters, defined for the small system, also determine the dynamics of large systems if
supplemented by the coupling constant $q$.

The CLM can be interpreted as a minimum approach to incorporate the dynamic coupling effects which
is compatible with the key observation that the thermodynamics as well as the diffusivity only show
very small finite size effect. 
On a qualitative level our model resembles the facilitation
approach since immobile regions are typically rendered mobile by the properties of nearby regions
and the elementary system size is temperature independent.
There is, however, one important difference. Whereas in the facilitation approach the
elementary systems are just the spins and therefore do not contain any relevant thermodynamic or
dynamic information, the elementary building blocks in the present case are small systems which
already contain the (nearly) complete information about the thermodynamics and the diffusivity.

In the literature one finds various arguments explaining the observed finite size effect that increasing the system size leads to a decrease of the relaxation time.
However, in a comprehensive study Berthier \emph{et al.} showed that this behavior is not   universal \cite{Berthier2012}. 
If the relaxation time increases with increasing system size this has been related to an activated mechanism (defect diffusion). In contrast, the opposite behavior has been explained in terms of a mode-coupling like mechanisms where the cooperative relaxation occurs via unstable modes. Going to small systems unstable modes may disappear. This explanation is very different to the present PEL approach because here the relaxation mechanism is described by a trapping-type picture rather than by unstable modes.

Karmakar \emph{et al.} found a remarkable correlation of the relaxation time with the configurational entropy \cite{Karmakar2009}, supporting the mosaic approach.
The data suggest that a system containing approx. 1000 particles can serve as unit system in terms of the configurational entropy which is much larger than the building blocks in the CLM. 
However, it remains open why already for much smaller systems sizes the diffusivity displays macroscopic behavior.
In \cite{Karmakar2012} the authors explain the finite size scaling behavior of $\tau_\alpha$ with the existence of a static length scale $\xi(T)$ and an entropic argument:
If the system size $L$ is smaller than $\xi(T)$ relaxation processes have to occur on the scale $L$ with a smaller degeneracy factor.
In larger systems the degeneracy factor will grow, leading to a lower free energy barrier.
In spirit this argument resembles the mode-coupling mechanism from \cite{Berthier2012}.
For the Kob-Andersen system $\xi(T)$ changes by roughly 30\% within the chosen temperature interval and our data suggest that a minimum system with around 65 particles is large enough to reproduce the thermodynamics as well as the diffusivity in this temperature interval.
Since the minimum system does not show a significant temperature dependence, we can only speculate that the relevant static length scales for all temperatures are already captured by the minimum system.

This entropy-related argument, and the CLM differ in one important point: While the earlier one is based on the free energy barrier height at a \emph{fixed time}, the finite size effect in the CLM is a consequence of fluctuating barriers (and corresponding jump rates) over a certain \emph{time interval} induced by the facilitation.
Since the free barrier from \cite{Karmakar2012} also determines the self diffusion it remains unclear how the different scaling behavior of $D$ and $\tau_\alpha$ can be reconciled with this approach.
In the CLM, at a fixed time the distribution of rates agrees with the equilibrium distribution of the minimum system. From this distribution directly follows the lack of finite size effects for the thermodynamics and the diffusivity, while the fluctuations effect $\tau_\alpha$ exclusively.

It may be interesting to study in future work whether it is also possible to specifically describe
the viscosity of large systems via a coupling of small systems. The observed strong correlations
between structural relaxation time and viscosity suggest that a similar approach should be
possible.

\begin{acknowledgments}
We appreciate the financial support by the DFG (SFB 458 and FOR1394) and we gratefully acknowledge the fruitful discussions and the important input from C. F. Schroer and O. Rubner. We thank L. Berthier for his helpful comments and discussions.
\end{acknowledgments}

\appendix
\section{Analytical results}\label{app: analytic}
Even in the generalized case it is possible to calculate
$\langle\Gamma^n(e)\rangle=\int\text{d}e\,p_{eq}(e)\Gamma^n(e)$ analytically as needed for the
calculation of $D$ and $\tau_\alpha$. Solving the Gaussian integrals one obtains
\begin{equation}
\begin{split}
\ln\average{\left(\frac{\Gamma}{\Gamma_0}\right)^n} =& \frac{n}{2}\lambda\sigma^2 \Bigg[
(n\lambda-2)\beta^2 + k_\text{entro}^2\kappa(n\lambda\kappa-2) \dots\\
&\dots+ 2\beta\left(k_\text{entro}(\kappa(1-n\lambda)+1)+\frac{V}{\lambda\sigma^2}\right) \Bigg ],
\end{split}
\end{equation}
the transport coefficients then read
\begin{equation}
\begin{split}
\ln D\propto&\average{\Gamma} =
\frac{1}{2}\lambda\sigma^2(\beta-k_\text{entro}\kappa)[\beta(\lambda-2)
-k_\text{entro}(\lambda\kappa-2)]\\
\ln \tau_{\alpha}\propto&\average{\frac{1}{\Gamma}} =
\frac{1}{2}\lambda\sigma^2(\beta-k_\text{entro}\kappa)[\beta(\lambda+2)
-k_\text{entro}(\lambda\kappa+2)]
\end{split}
\end{equation}

\begin{figure}[tb]
\includegraphics[width=\picturewidth]
{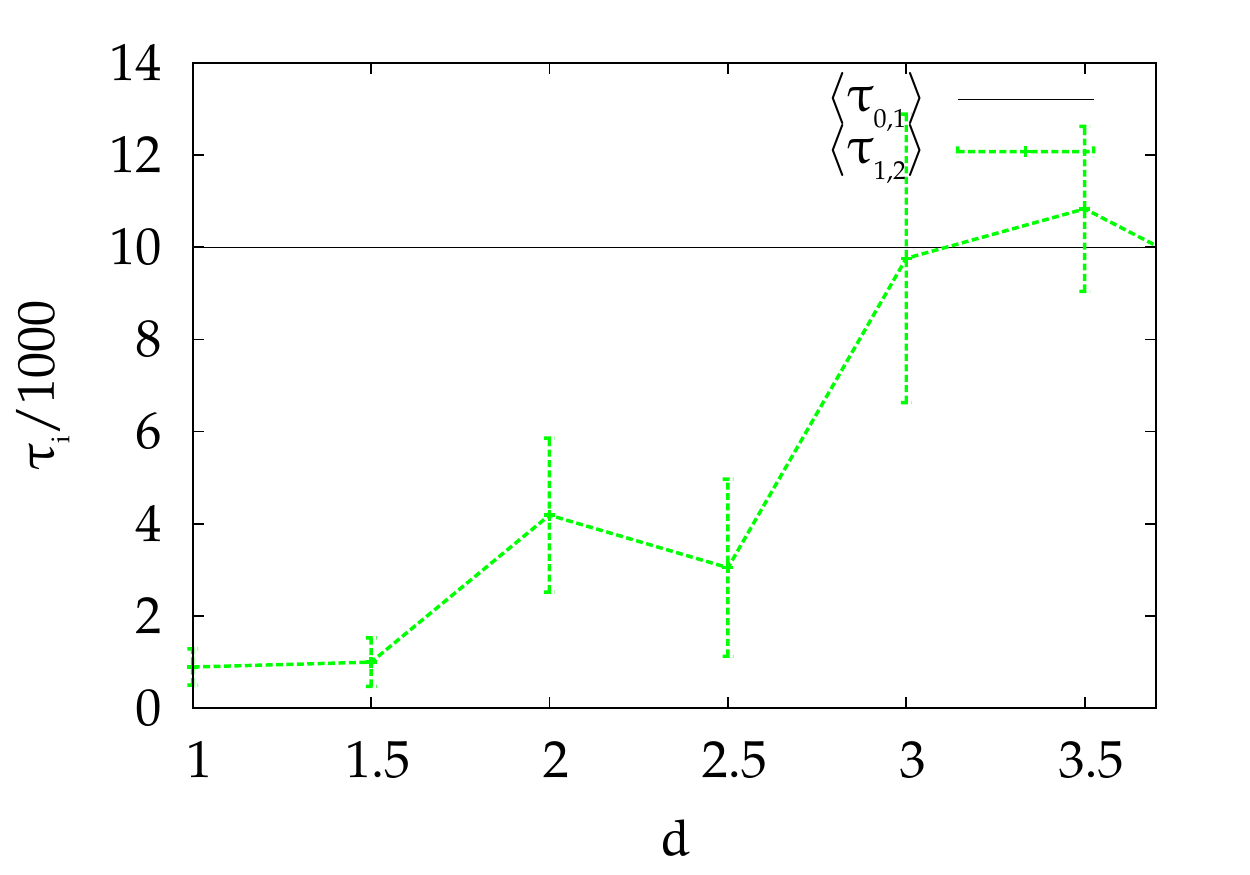}
\caption{The different waiting times (for the definition see text) for a $MSD$ threshold of 2 vs.
the corresponding distance.}
\label{fig: MSD non-equi}
\end{figure}

\section{How to handle a system with \texorpdfstring{$\lambda<1$}{lambda<1}}
When modeling a BMLJ65 with e.g. two independent PEL systems one faces a problem:
In the MD one can only access $\Gamma_2(e)$. If one calculates the entire rate $\Gamma_N(e)$ of
two PEL systems with $\Gamma_1(e)$ for the model system from \cite{Heuer2008}, the superposition
leads to an additional temperature dependent factor $F(T)$ (from the thermodynamic distribution of
energies). In the general case it is even possible to end up in an additional energy dependence,
making it impossible to extract $\Gamma_1(e)$ from the MD data.

When one uses a single system with $\lambda<1$ one hast to take into account that the persistence
time distribution is no longer exponential and has to be determined numerically when simulating the
system. Furthermore, $\average{\tau^2(e)}/\average{\tau(e)}^2>2$ as obtained in the MD
simulation  has to be included separately by the distribution $\varpi(\gamma,e)$ (see text). Another
result is that the invariance of $\average{\tau}$ under passive processes is no longer valid since
$\average{\tau(e)}=1/\average{\Gamma(e)}$ does not hold for $\sigma_\Gamma>0$. However, this
effect is very small and even for high coupling strength smaller than $\approx$ 4\% (at $T=0.45$).

\section{Waiting times of the non-equilibrium ICE} \label{sec: tau12}
As discussed in the text a possibility to gain information about the rate fluctuations of
the coupling is the waiting time  $\average{\tau_{1,2}}(d)$ between the first two structural events
in
dependence on their distance $d$. In the case that subsequent relaxation processes are uncorrelated
or the coupling mechanism uses a fixed rate one would expect
$\average{\tau_{1,2}}(d)=\average{\tau_{0,1}}$. In the presence of coupling with a rate
distribution, small $d$ will display small $\average{\tau_{1,2}}(d)$. The results are shown
in Fig. \ref{fig: MSD non-equi}. We found $\average{\tau_{1,2}}$ increases monotonically with
growing distance $d$ up to half of the cell length. Very small values of $\average{\tau_{1,2}}$ are
seen for
$d<2.5$, followed by a large jump in $\average{\tau_{1,2}}$. These findings again contradict the
statistical case, but since we are in a non-equilibrium configuration we cannot extract information
about the strength of the rate fluctuations. The waiting time $\tau_{0,1}$ for the first
structural event is also included in the figure. The matching of both observables at large $d$ is
to our knowledge highly nontrivial.

%Merlin.mbs v4.21 2009-07-09.
%

% \bibliographystyle{aipnum4-1}%plaindin,plainnat,plain,aipnum4-1,aipauth4-1,unsrtnat,
% \bibliography{../../Dissertation/bib/general_pre,../../Dissertation/bib/kcmModels_pre}

\end{document}